# Striking Isotope Effect on the Metallization Phase Lines of Liquid Hydrogen and Deuterium

Mohamed Zaghoo, Rachel Husband, and Isaac F. Silvera*

Lyman laboratory of Physics, Harvard University, Cambridge, MA 02138

**Liquid atomic metallic hydrogen is the simplest, lightest, and most abundant of all liquid metals[1,2]. The role of nucleon motions or ion dynamics has been somewhat ignored in relation to the dissociative insulator-metal transition. Almost all previous experimental high-pressure studies have treated the fluid isotopes, hydrogen and deuterium, with no distinction[3-8]. Studying both hydrogen and deuterium at the same density, most crucially at the phase transition line, can experimentally reveal the importance of ion dynamics. We use static compression to study the optical properties of dense deuterium in the pressure region of 1.2-1.7 Mbar and measured temperatures up to ~3000 K. We observe an abrupt increase in reflectance, consistent with dissociation-induced metallization, at the transition. Here we show that at the same pressure (density) for the two isotopes, the phase line of this transition reveals a prominent isotopic shift, ~700 K. This shift is lower than the isotopic difference in the free-molecule dissociation energies[9], but it is still large considering the high density of the liquid and the complex many-body effects. Our work reveals the importance of quantum nuclear effects in describing the metallization transition and conduction properties in dense hydrogen systems at conditions of giant planetary interiors, and provides an invaluable benchmark for ab-initio calculations.**

As the lightest atoms, hydrogen and its isotopes exhibit the largest mass ratios of the elements, giving rise to large differences in their properties. The binding energies of the homonuclear free-molecules $H_2$ and $D_2$ differ by ~900 K (4.477eV for $H_2$ and 4.556 eV for $D_2$)[9]. The isotopic shift in the binding energy is related to the different zero-point energies (ZPE) arising from the fundamental vibrational mode of the molecules. At low temperatures (T), where both hydrogen and deuterium form quantum solids due to their ZPE, the large isotopic effects are manifested in phonon, vibrational, and rotational excitations, as well as differences in their equations of state and melting temperatures[10]. In 1935, Wigner and Huntington first discussed the role of density in destabilizing the molecular bond in the crystalline phase, predicting a dissociative transition to metallic hydrogen (MH) at sufficient compression[11]. This transition has recently been reported in hydrogen at low temperatures by Dias and Silvera[12]. In the high-temperature, high-density fluid phase, a liquid-liquid insulator to metal transition (LLIMT) has long been studied to describe molecular hydrogen's transition to atomic liquid metallic hydrogen (LMH)[13,14], also called the plasma phase transition (PPT)[15-22]. This LLIMT locates the boundary between a molecular fluid and atomic MH in astrophysical gas giants, a central characteristic of all interior model structures of Jovian-like planets[2]. Due to its fundamental nature as well as its astrophysical significance, the precise location and nature of this phase line remains the subject of intense theoretical and experimental research.

Early theoretical studies differed in their predictions of the LLIMT boundary by as much as 4 Mbar in pressure (P) and ~10,000 K in T. All theories predict the transition to be first-order with a negative P,T slope and a critical point at the low P, high T side of the line. However, early studies that were focused on hydrogen did not explicitly consider the heavier isotopes. Although many theoretical studies treat the protons classically in the warm dense fluid[16,18], the inclusion of nuclear quantum effects can significantly alter the location of the phase line (see discussions in ref.[23]). In a recent study the phase lines for isotopes were calculated using coupled electron-ion Monte Carlo (CEIMC)[20], which treats the ions quantum mechanically (Fig. 1). Results were in excellent agreement with the phase line of hydrogen determined by static compression techniques[23], but indicated a small isotopic shift in the phase line of order ~ 20-40 GPa for deuterium, or ~100-200 K. Here we report the observation of a considerably larger difference in the metallization temperatures of the isotopes. We attribute this difference to the isotopes' different dissociation energies.

There are two experimental techniques that have been used to study the high-temperature metallization of the hydrogen isotopes: 1) dynamic compression techniques that can pressurize and heat the sample for a few to tens of nanoseconds, depending on the technique, and 2) diamond anvil cells (DACs) that produce high static pressures and high temperatures using CW or pulsed laser heating that lasts a few hundred nanoseconds. Single-shock Hugoniots heat the sample to very high measured temperatures (~$10^4$ K or higher)[4,8] and can arrive in the metallic phase, as determined by reflectance of the sample, but have not detected the phase line. In dynamic experiments designed for lower temperatures, the temperature is often estimated by calculation. The thermodynamic pathway can either cross the first-order phase line or can enter the metallic phase below the critical pressure or above the critical temperature where the transition is continuous. Weir et al[3] used reverberating shocks to probe a lower temperature region and observed metallization of $H_2$ and $D_2$ by the change in conductivity. They did not determine the phase line. They point out that gas gun experiments are not designed to probe isotope effects; the initial mass densities differ by 2.4 so their final molar densities and temperatures differ substantially at the same final pressure[24]. Knudson et al[6] used ramped magnetic compression of $D_2$ to probe lower temperatures. They measured pressure and calculated temperatures, reporting the sample to first become absorptive and then reflective, and presented a phase line based on the abrupt increase of reflectance (Fig. 1). Fortov[5] and Mochalov[7] and colleagues also studied the density discontinuities, using explosive compression of deuterium. There is a clear discrepancy between the measurements of Knudson et al and the other dynamic compression experiments on deuterium.

The location of the phase line of hydrogen was determined in DACs by Dzyabura et al[25] and Ohta et al[26], based on documented plateaus in heating curves (plots of peak temperature vs. laser power). Zaghoo, Salamat and Silvera[23] and Zaghoo and Silvera[27] studied the optical properties of hydrogen at static pressure conditions, showing that the observed transition was to liquid atomic metallic hydrogen (Fig. 1). The transition was identified by absorption of light that coincided with the onset of the plateaus, while the rise of reflectance was observed for temperatures above the plateau temperature.

In order to clarify the discrepancy with deuterium we have measured the transition line to the liquid metallic state of deuterium using the same experimental method that was previously used for hydrogen[23,27]. This enables us to make a comparison while eliminating possible systematic uncertainties associated with comparison of data collected using different techniques. We have studied the optical properties of statically compressed deuterium samples in the pressure range of 124-169 GPa and at temperatures up to ~3000 K using pulsed laser heating (see Methods and Supplementary Information, SI). In order to better compare with the majority of dynamic studies that use reflectance as the signature of metallization, in this paper we use the onset of reflectance as the criterion for the phase line. Figure 1 shows our P-T points for the onset of the highly reflecting/conducting state in liquid deuterium and can be compared to results for hydrogen. In Fig. 2 we show the reflectance as a function of time for a series of increasing temperatures; in Fig. 3 we show the reflectance as a function of temperature at three pressures.

Optical reflectance data are amenable to the Drude free-electron analysis, which has two parameters: the plasma frequency $\omega_p$, where $\omega_p^2 = 4\pi e^2 n_e/m_e$, and the scattering time $\tau$. Here $n_e$ and $m_e$ are the electron density and mass. The measured reflectance for a thick metallic film at a given frequency is $R(\omega) = |(N_{diam} - N_D)/(N_{diam} + N_D)|^2$, where $N_{diam}$ is the index of refraction of the molecular deuterium/diamond layer and $N_D$ is the complex index of refraction of LMD: $N_D^2 = 1 - \omega_p^2/(\omega^2 + j\omega/\tau)$. We performed a least-squares non-linear fit of our measured R(ω) to extract these parameters (Fig. 3, lower right panel). The highest values of our measured reflectances for LMD (Fig. 3) are around 0.5-0.6, overlapping the values found for reflectance in dynamic experiments[4,6,8] and thick hydrogen films. As the pressure or temperature increase beyond the onset region, LMD becomes Drude–like and when the carriers are degenerate, the free-electron character is established. At 142 GPa and 2442 K, the data are best fit to ω$_p$ =21.3±0.9 eV and τ of 1.7±0.2 x10$^{-16}$ s; the dissociation fraction is 0.70±0.06, calculated from the plasma frequency and the atomic density[28].

There have been two models proposed for the IMT: dissociation and electronic band overlap[29]. Band overlap is predicted to take place at the same pressure for both isotopes; such a mechanism is eliminated by the large observed isotopic differences. The shift of the metallization phase line must be due to the isotopic difference in the dissociation energies in the dense fluid phase. A negative slope of the metallization phase line, observed in our data, is also supportive of the dissociation model; the dissociation energy decreases with density and thus the transition temperature decreases.

Although our current data does not extend to high enough pressures to make a P-T comparison isothermally (Fig. 1), at the same pressure we find an isotopic shift on the order of ~700 K, attributed to differences in the molecular dissociation energies due to ZPE. The magnitude of the shift is consistent with the difference in the ZPE of the vibrational energies of the two species at high density (see statistical estimate in SI). This is striking, considering the complex many-body effects, strong coupling, and thermal degeneracy present in the dense fluid state. Our hydrogen metallization phase line (based on the onset of reflectance) is in rather good agreement with the most recent theoretical results[20,22], however the CEIMC calculation[20] appears to underestimate the observed isotope shift.

Apart from revealing a fundamental effect in the metallization phase line in the dense hydrogen isotopes, our results with measured temperatures could help to resolve the conflicting results in dynamic experiments. As shown in Fig. 1, our deuterium phase line is in good agreement with earlier shockwave studies on deuterium[3,5,7], albeit their large temperature uncertainties. Our data is also consistent with the new laser compression experiments at the National Ignition Facility, which identified metallization in compressed deuterium at 200 GPa and above 1000 K[30]. If the calculated temperatures by Knudson et al[6] were scaled to lower values, their data might fall onto an extrapolation of our deuterium phase line. Modeling of interior planetary structures of Jovian-like planets, where hydrogen (not deuterium) is the primary constituent, should thus rely on the previous static pressure[23,27] determination of the PPT phase line.

The most recent magnetic measurements from the Juno space mission are also supportive of this conclusion, where the Jovian dynamo action is revealed to operate at much shallower depths than previously estimated[31], consistent with our hydrogen metallization line reported at lower pressures. Furthermore, our work suggests the importance of nuclear quantum effects in describing the electronic and thermodynamic states of planetary interior hydrogen-rich dense warm ices, such as water, methane or ammonia. If so, then current thermal and magnetic models of ice-giants[32], which have thus far relied on ab-initio DFT methods implementing classical ions, may require revision.

## Methods

The heart of the diamond anvil cell (DAC) used to study deuterium is shown in the lower panel of Extended Fig. 1. Pressurized deuterium was pulse-laser heated at a 20 kHz repetition rate using thin tungsten (W) films coated on a diamond as absorber. The W absorbs the Nd-Yag heating laser light (1064 nm) and is heated, which in turn heats the deuterium pressed to its surface. The W also reflects and transmits light from CW lasers with wavelengths in the visible and near infrared region. Time-resolved spectroscopy was used to measure the optical properties at these wavelengths. By incrementally increasing the laser power or energy/pulse of our heating laser, we gradually increase the peak temperature of the deuterium sample. When the deuterium is sufficiently hot, we observe an abrupt increase in optical reflectance for all the frequencies investigated, shown in Fig. 2. At the threshold of the transition, the liquid metallic deuterium (LMD) film is thin so that it is partially transparent, but as the laser power is increased the film thickens and reflectance signal rapidly increases.

**Pressure and Temperature Determination**

Pressure was determined from Raman scattering of the deuterium fundamental vibron $Q_1$ mode at room temperature, with an uncertainty of ±2 GPa (see SI). An advantage of the work reported here over our earlier studies on hydrogen is that the lower energy deuterium vibron shown in the upper panel of Extended Fig. 1 falls within the spectral range of our InGaAs diode array detector. This enabled us to monitor the vibron, and thus the integrity of our deuterium samples as a function of increasing temperature shown in Extended Fig. 2. Pressurized samples were isochorically heated to temperatures as high as 3000 K and temperature was determined from fitting the emitted thermal irradiance spectra to a Planckian curve. Examples of the collected raw data and the fitting region are shown in Extended Fig. 1. For the temperature fit, the Raman spectra were edited out, using a procedure described elsewhere[25] (Extended Fig. 1 and SI).

**Reflectance Measurements**

Time-resolved reflectance of hot dense deuterium samples was measured at two distinct wavelengths: 514 nm and 980 nm. Fast Si-detectors, triggered by the heating laser, recorded the reflected light intensity while temperature was measured. Unlike all dynamic studies of the PPT, the photon energy dependence of optical reflectance was measured simultaneously with measurements of the temperature. Examples of the reflectance data as a function of time are shown in Fig. 2, for increasing temperature at 142 GPa. For times before the heating laser pulse, the trace shows the reflectance signal from the W absorber surface, $R_W$. Below the transition temperature, no detectable change in $R_W$ is detected. Above the transition, a sharp rise in reflectance is observed due to the transition to liquid metallic deuterium, consistent with a first-order phase transition. We normalized the intensities of the traces with reflective states (above the transition) to those below the transition to determine the magnitude of metallic deuterium reflectance. The analysis of the reflectance data is identical to the analysis performed on hydrogen where a detailed description of the procedure is given[23]. At the relevant pressures the real index of diamond and molecular deuterium are very close in value so that they index match and thus a multilayer analysis is not needed, as discussed elsewhere;[27] small differences in the indices do not affect the results. We contrast our optical reflectance and the Drude fit to that found for LMD using the minimum relaxation times, $\tau_{min}$, prescribed by the Mott-Ioffe-Regal (MIR) minimum metallic conductivity limit (electron scattering length is the average spacing between particles), assuming full dissociation. As shown in Fig. 3, our measured reflectances for LMD are much higher than that expected for the MIR limit.

The goal of the analysis is to isolate the reflectance of the hydrogen from that of the tungsten. We reproduce an important result of that analysis, relating the reflectance R to the reflectance signal Rs: $R = R_W[R_S - (Tr)^2]$. Here Rw is the reflectance of the tungsten and Tr is the transmission of the metallic film. For thick films R=RwRs, since the transmission approaches zero. Our reflectance data are summarized in Fig. 3 and compared to work on hydrogen[27]. The value of Rw for a W film deposited on diamond was measured in a few cases. For each deposition of W on a diamond, we co-deposited on a glass slide and measured the transmission of the slides to create a correlation table between optical properties of the coated slides and the coated diamonds. We used these correlations to determine the values of Rw shown in Extended Table I, for different runs where the thickness of the W was different. Measured values of Rw were found to be in good agreement with results reported in the table. We estimate the uncertainty in Rw to be 10%, using this method.

Acknowledgements: The NSF, grant DMR-1308641, the DOE Stockpile Stewardship Academic Alliance Program, grant DE-NA0001990, and NASA Earth and Space Science Fellowship Program, Award NNX14AP17H supported this research. Preparation of diamond surfaces was performed in part at the Center for Nanoscale Systems (CNS), a member of the National Nanotechnology Infrastructure Network (NNIN), which is supported by the National Science Foundation under NSF award no. ECS-0335765. CNS is part of Harvard University.

Author Contributions: all authors contributed to all aspects of this letter: design, measurement, analysis, and writing. Data is available from the authors upon request.

8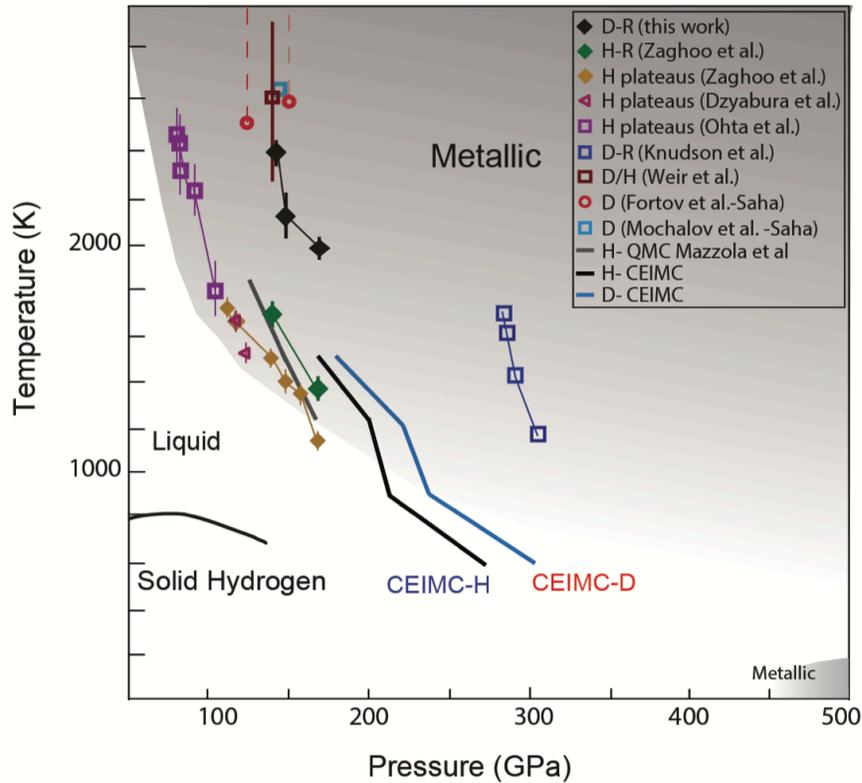

**Fig. 1**. The hydrogen-deuterium phase diagram showing experimental data and theoretical predictions of the insulator-metal phase boundary in the fluid phase, as well as the theoretical melting line for hydrogen, all identified in the legend. Our data points (solid black diamonds) delineating a phase boundary are based on the optical signature of an abrupt rise of reflectance. Systematic uncertainties in the temperature are discussed elsewhere (SI). The green diamonds indicate the onset of reflectance from our previous hydrogen studies, based on reflectance; we also show the phase line based on observation of plateaus[23,26]. The P-T values where static experiments reported metallization at low temperatures are also shown[12]. The solid black and blue lines with circles indicate theoretical CEIMC predictions for the PPT phase lines in hydrogen and deuterium, respectively[20].

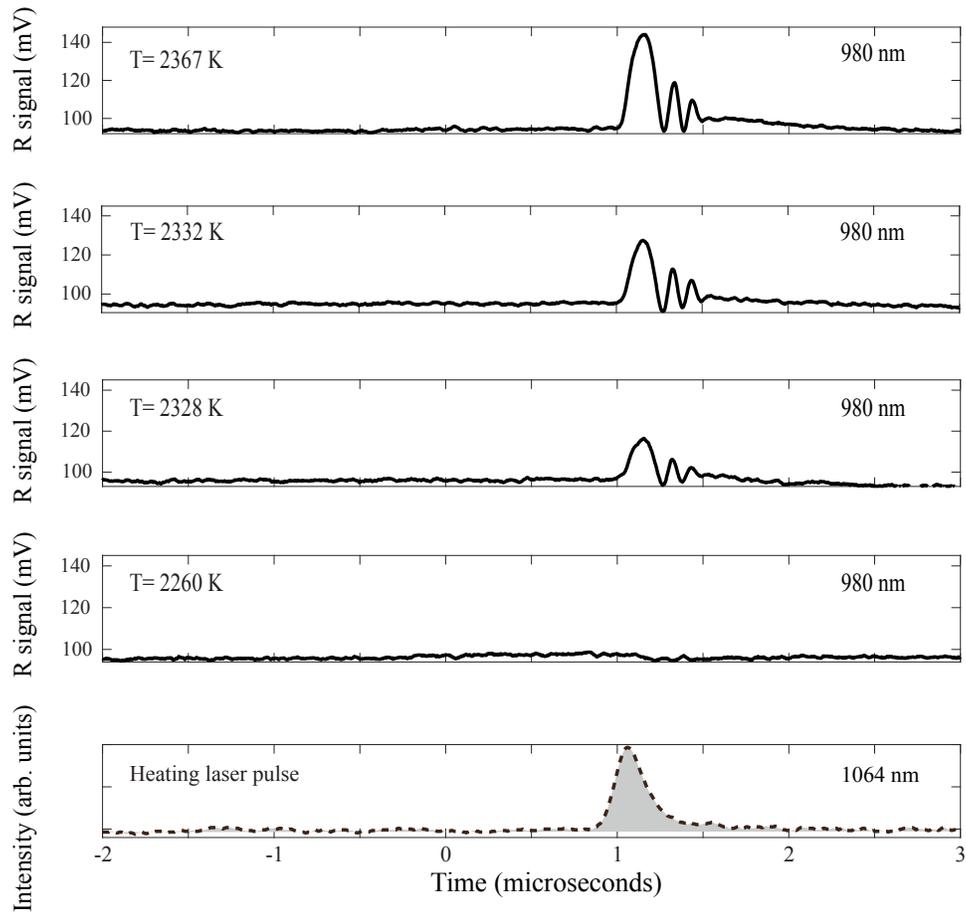

**Fig. 2**. Example of the raw time-resolved reflectance traces recorded as a function of temperatures and time for 980 nm light and P=142 GPa. The bottom panel shows the timing of the heating laser-pulse, 1064 nm. Above a certain temperature, liquid deuterium abruptly metallizes and reflects the incident probe light. Increased heating thickens the metallic layer that reflects more light. The secondary peaks are due to interference in the metallic deuterium/diamond interfaces.

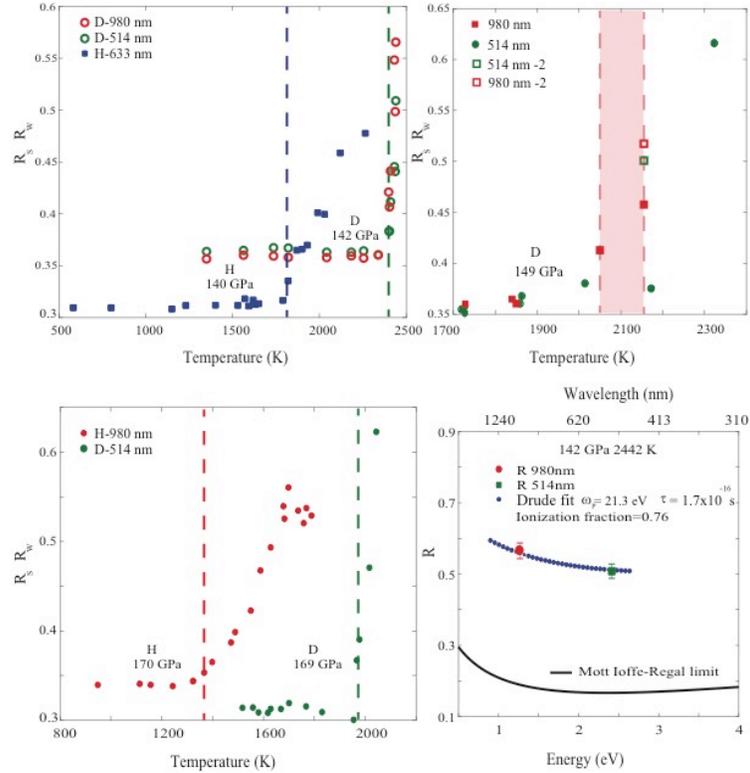

**Fig. 3**. Top and bottom left: reflectance signal of deuterium and hydrogen on a tungsten film versus T for various pressures. Both isotopes reflect more light at increasing wavelengths, consistent with the free-electron model. Hydrogen data are from ref.[27]. Open symbols represent data in which reflectance was measured simultaneously at the two wavelengths. For the 149 GPa data (top right), the open symbols denote a different optical run than the filled ones. We plot the reflectance signal times the reflectance of the tungsten film. This is equal to the metallic reflectance for thick films (low transmittance, see Methods). The abscissas are a measure of the reflectance of W below the transition, and the reflectance of liquid metals for thick films. For the deuterium we estimate a possible systematic uncertainty of 10% for the reflectance of the tungsten film; this is larger than the random errors so we do not show error bars; random temperature uncertainties are small, discussed in the SI. Bottom right: the Drude fit to the experimental data derived from a least-squares fit to the energy dependence of measured reflectance. The solid line is the calculated reflectance using the Mott-Ioffe-Regal (MIR) limit, assuming full dissociation. The vertical lines indicate the T values for the transition, based on reflectance.

# Supplementary Information

## Striking Isotope Effect on the Metallization Phase Lines of Liquid Hydrogen and Deuterium

Mohamed Zaghoo, Rachel Husband, and Isaac F. Silvera

Lyman laboratory of Physics, Harvard University, Cambridge, MA 02138

**Summary of experimental work**

A total of 8 deuterium samples were loaded in two different DACs. Optical data were collected on 4 samples (details are in Table I). The other 4 samples suffered diamond breakage on pressure increase between 142 and 169 GPa before any reflectance data were collected. We used either type Ia or type IIac beveled diamonds with 100/300 μm tips (culet/bevel diameter); alumina coated rhenium gaskets were used in all cases. The initial height of the sample chamber in the gasket was ~10 μm and the diameter ranged from 55-65 μm. The prepared gasket was epoxied onto the bottom diamond and then the diamonds were coated with ~50 nm amorphous alumina ($Al_2O_3$) using atomic-layer deposition (ALD). Alumina inhibits hydrogen diffusion into the diamond anvils; diamond embrittlement from hydrogen diffusion is an important source of diamond failure in high-pressure DAC experiments. A thin, semi-transparent tungsten (W) film, ~8.5-11 nm, was deposited onto the top diamond using sputter-deposition, and then coated with a ~5 nm amorphous $Al_2O_3$ protective layer to inhibit hydrogen diffusion into the W.

| Sample | Pressures at which data were collected (GPa) | Diamond type | Maximum temperature during heating (K) |
|---|---|---|---|
| 1 | 142 | IIac, CVD | 1770 |
| 2 | 156 | Ia | 1534 |
|   | 169 |    | 2032 |
| 3 | 124 | Ia | 1920 |
| 4 | 142 | Ia | 3092 |
|   | 149 |    | 2353 |
| 5 | 146 | Ia | --- |
| 6 | 169 | Ia | -- |
| 7 | 142 | Supplier: Washington diamonds, CVD | -- |
| 8 | 153 | IIac, CVD | -- |

Table SI I. The optical data described in the main paper were all collected from samples 2 and 4. In samples 1 and 3, the sample was not heated to a high enough temperature to observe the transition to LMD. CVD is chemical vapor deposition synthetic diamonds.

A schematic of the interior of the DAC is shown in Extended Figure 1. Deuterium samples (99.9% pure) were cryogenically loaded into the DAC and warmed to room temperature for further study. The sample was compressed to the desired pressure at room temperature. At high pressure, the deuterium sample was ~20-30 μm in diameter. Heating of the sample and the optical measurements were performed using the optical set-up described in refs.[1,2]

**Temperature determination**

The sample was pulse-laser heated using a Nd-YAG laser (wavelength 1064 nm) at a 20 kHz repetition rate. Pulse-power was varied incrementally using a polarizing cube and half-wave plate so that the shape of the laser pulse remained the same for all pulses (see SI of Ref.[1]). The pulse width was 290 ns which is long enough to achieve local thermal equilibrium in the heated deuterium, but sufficiently short to suppress hydrogen diffusion. The energy per pulse ranged from ~0.007-0.07 mJ. The laser couples to the W absorber and the hydrogen pressed against the W is heated by contact. The W film is much thinner than the optical attenuation length, so it is uniformly heated and thermalizes in the order of several picoseconds. In this region the measured blackbody (BB) radiation is from the W absorber. At temperatures above the transition, the deuterium pressed against the heated absorber is transformed to LMD.

BB radiation from the heated sample was collected by a Schwarzschild objective, magnified, and imaged onto a 200 μm pinhole to separate contributions from the sample and the surrounding gasket. Light was dispersed using a prism spectrometer and detected by an InGaAs diode array. Collection times ranged from 50 s at low laser power to 180 ms at the highest laser powers; the signal was averaged over many heating pulses. The collected spectra were corrected using a transfer function that accounts for the optical response of the system. This was determined by measuring a spectrum from a W-coated diamond anvil that was ohmically-heated to 800 K which enables the wavelength dependence of the emissivity of the W absorber to be included in the transfer function[3]. Background contributions corresponding to Raman scattering from the deuterium sample and the diamond anvils, and fluorescence from the diamond anvils was removed using methods described in ref.[4]. Data processing is illustrated in Fig. SI1. The transfer function is also shown along with example spectra before and after correction.

Each spectrum consists of BB contributions averaged over the entire laser pulse. An average temperature is first calculated and then converted to a peak temperature using the method outlined by Rekhi, Tempere, and Silvera[5] and Deemyad et al.[6]. The Planckian average temperature was determined using two methods. First, by fitting the corrected spectra based on Planck's law (eq. 1). Second, by fitting to a linearized form of Planck's law, the Wien approximation, which assumes that $hc/\lambda \gg k_b T$. In Eq. 2 below $I_{BB}$ corresponds to the photon flux. The same wavelength region (1121 to 1609 nm) was used to determine the temperature in all cases. The temperatures determined from the Wien and Planck fits generally differed by several degrees, and if the difference was significant we used eq. (1).

$$I_{BB} = \frac{2ec}{\lambda^4} \frac{1}{\exp(hc/\lambda k_v T) - 1} \quad (1)$$

$$\log(I_{BB}\lambda^4) = \log(2ec) - hc/\lambda k_b T \quad (2)$$

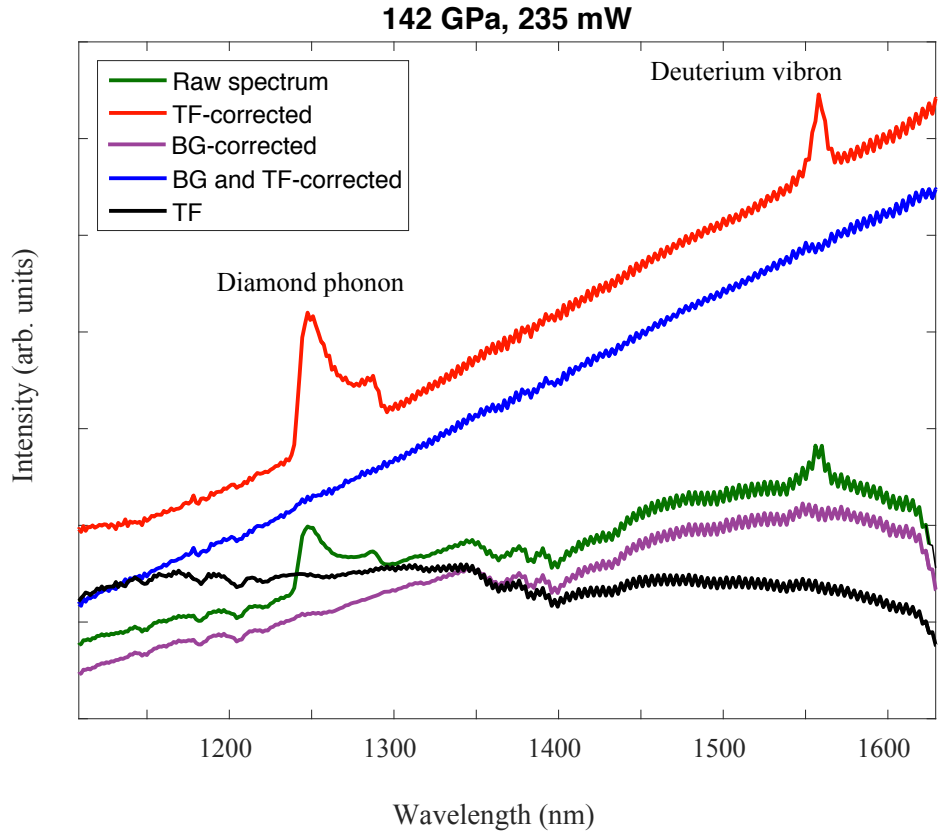

Figure SI1. An example of the correction process to the spectra collected on the diode array. Spectra are corrected for the background (BG) and for the transfer function (TF). In this figure the TF is multiplied by 50,000 and the TF-corrected spectra are divided by 8, to allow for comparison on the same figure. Spectra were collected at 142 GPa using 235 mW average laser power. The BG and TF corrected is the BB spectrum used to determine the temperature. Oscillations at longer wavelengths are due to etaloning arising from optical interference in the diode array.

Examples of the Planck and Wien fits based on the data presented in Fig. SI1 are shown in Fig. SI2. The calculated temperatures are 1509(2) and 1527(4) K, respectively, where the uncertainty is from a least-squares fit. Both give excellent fits. Using a correction table[5], yields peak temperatures of 1689(2) K and 1712(4) K.

Heating curves (peak temperature against average laser power) for optical data reported in Fig. 4 of the paper are shown in Fig. SI3. Both the Planck and Wien fits are shown for comparison. The uncertainties from the fits are small (less than 15 K) for all except for the very lowest temperatures which are less important; the true uncertainty can be larger due to systematic changes such as pressure dependence of the emissivity. Furthermore, spectra collected above the transition contain contributions from the LMD. Note that a thin metallic film matched to the impedance of free space absorbs 50% of the radiation, and thus the emissivity is 0.5. Bulk LMD also absorbs about 50% of the radiation so it also has an emissivity of ~0.5. We try to match the emissivity of the W to that of the LMD, to minimize uncertainties in the temperature determination above the transition temperature, as the wavelength dependence of the emissivity of the LMD is not known. This can add a systematic error to the temperature.

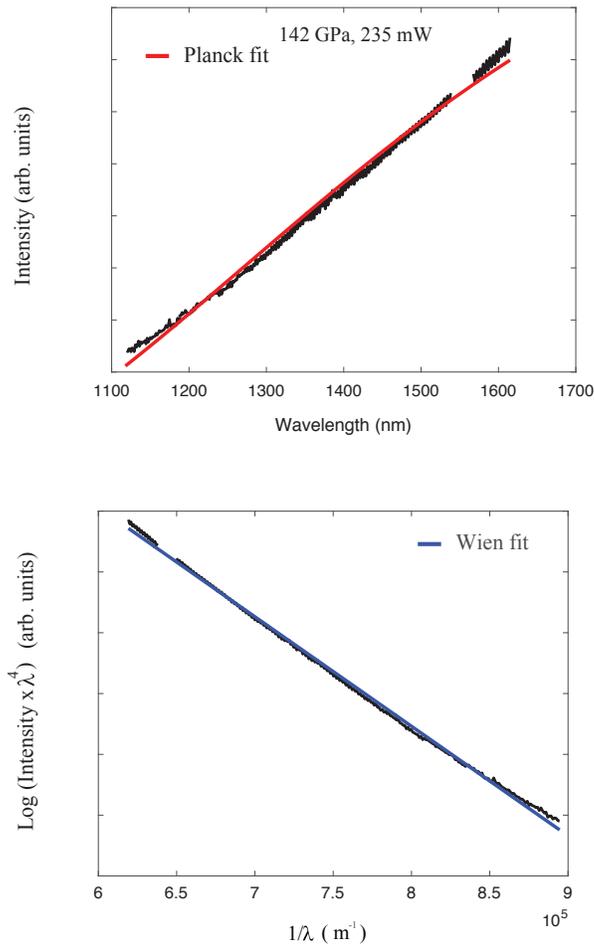

Figure SI2. Fit of the corrected BB spectrum in Fig. SI1, based on the Wien approximation and on Planck's law for the spectrum collected at 142 GPa with 235 mW average laser power.

5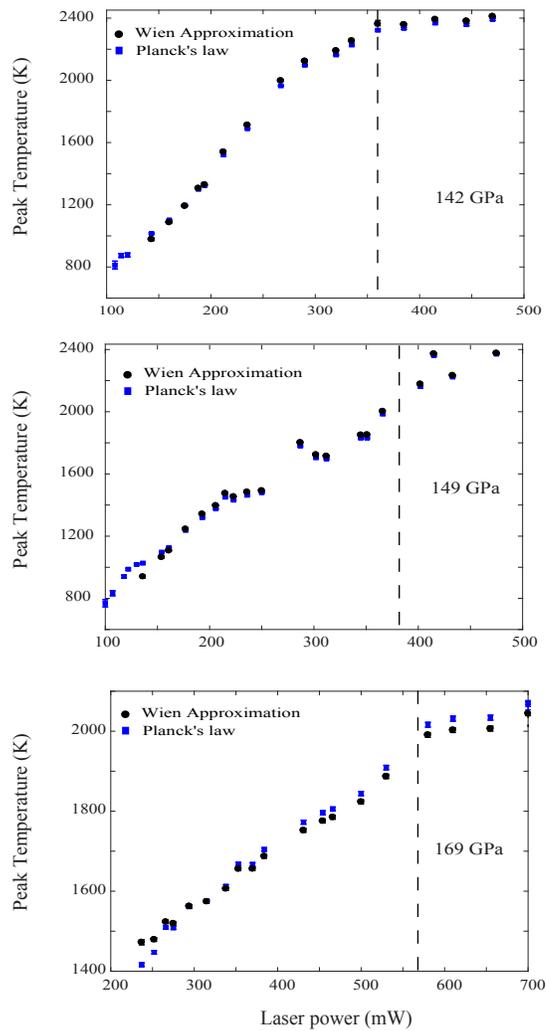

Figure SI3. Heating curves collected at 142, 149 and 169 GPa, where temperatures were determined using fits based on Planck's law and based on the Wien approximation. The dashed lines indicate temperatures above which an increase in reflectance was observed.

In spite of the small uncertainties of the fits, accurate estimation of the uncertainty in the temperature determination based on pyrometry measurements is challenging. Our transfer function correction significantly reduces the uncertainty associated with the wavelength dependence of the emissivity[3]. Nevertheless, we have performed a two-color analysis as outlined by Benedetti and Loubeyre[7] to determine a value for temperature and its uncertainty. This method was developed to quantify uncertainties due to temperature gradients, as well as the wavelength-dependent emissivity. We have applied this method to the heating curve collected at 142 GPa; the resulting uncertainties are shown in Fig. SI4. The largest uncertainty is of order ~125 K for all but the lowest temperatures. We believe that the two-color method may be more appropriate for CW heating, but exaggerates the uncertainty for our pulsed laser heating. For pulsed heating we average the spectral irradiance over the pulse shape to determine a Planckian temperature and then use a calculated look-up table to determine the peak temperature[5]. The uncertainties shown in Figs. 1 and SI7 thus reflect the compromise between uncertainties that are too small and those that are too large. These uncertainties should be compared to those in shock experiments where temperature is not measured and calculated uncertainties are thousands of degrees K.

In earlier work on hydrogen[1], the phase transition to LMH was determined by the observation of plateaus in the heating curves. Plateaus were documented at powers lower than the onset of reflectance. In those heating curves, increasing the laser power beyond values corresponding to plateaus, the temperature again rose and the LMH thickened so that the reflectance achieved bulk values. For the heating curves of deuterium, plateaus are difficult to identify, but we could achieve conditions such that the reflectance rose rapidly to an almost saturated value. However, we could not easily heat the sample to higher temperatures, as shown in Fig. SI3. Thermodynamic conditions for the heating curves are evidently quite different in this substantially higher temperature range.

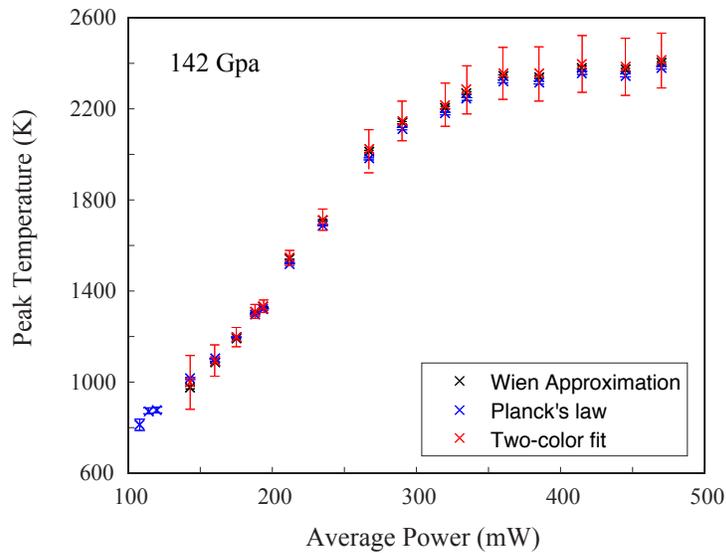

Figure SI4. Heating curve collected at 142 GPa, showing temperatures and the corresponding uncertainties, applying the two-color method. The average power refers to the sampled heating laser power, measured with a CW power meter. The actual power was 5 times this value.

**Pressure Determination**

The pressure of the sample was determined from the Raman shift of the deuterium vibron relating pressure to the peak frequency[8]. An advantage of our current work over earlier work on hydrogen[1] is that the Raman scattering by the deuterium vibron excited by the heating laser falls in the wavelength range of the diode array. The corresponding peak is observed in the BB spectrum shown in Fig. SI1. This enabled us to monitor the integrity of the deuterium sample at increasing temperatures. Additionally, we have regularly checked the vibron after the heating cycles. The region of the raw spectrum showing the deuterium vibron is displayed in Fig. SI5 for a range of temperatures. Observation of the vibron at temperatures above the transition temperature is not an indication that LMD is molecular in character. Only a small fraction of the deuterium is converted to LMD, and the remainder is still molecular. Sample thicknesses are ~2000 nm while the thickness of LMD films are believed to be 10-100 nm at high pressure. Moreover, the spectrum is collected over the entire heating pulse; the cell is isochoric so that the pressure rises with temperature. The molecular hydrogen is in a thermal gradient and Raman contributions come from this thermal distribution. The pressures reported are based on the known vibron peak frequencies at room temperatures. Uncertainties are ±2 GPa.

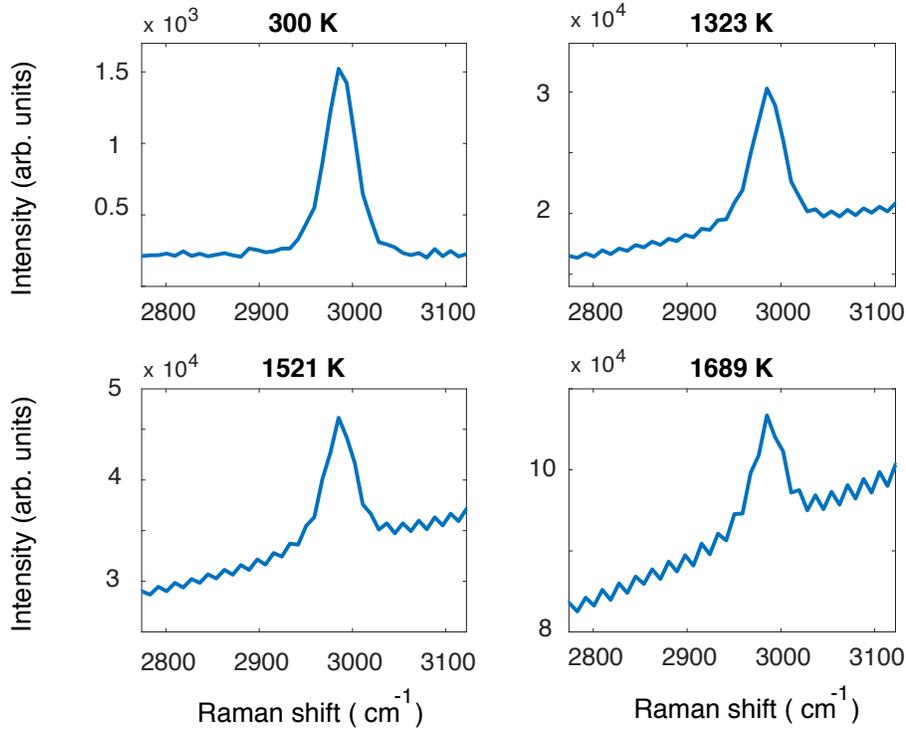

Figure SI5. Deuterium Raman vibron spectra over a range of temperatures, including room temperature, during a heating cycle at 142 GPa. The higher temperature spectra appear noisier as the vibron spectrum sits on top of the large blackbody background.

**Optical measurements**

In this work, we focused on reflectance measurements to allow for direct comparison with dynamic compression experiments such as those reported by Knudson et al.[9] Time-resolved reflectance measurements were conducted using CW solid-state lasers at two wavelengths, 514 and 980 nm, with power in the 5-10 mW range. The reflected signal was collected on Si photo-detectors with a 3 ns rise time, and monitored on a 155 MHz oscilloscope, triggered by the laser pulse. In order to compare reflectance at different photon energies they have to be measured at the same temperature due to the rapid rise of reflectance at the onset. As determined elsewhere[1] temperature uncertainty could be as large as ±50 K from heating cycle to heating cycle due to small mechanical instabilities. A method was developed for simultaneous measurement of reflectance at two wavelengths at a given temperature, described elsewhere[2]. Reflectance and transmittance could not be measured simultaneously. Transmittance is optimally measured for normal incidence to the diamond tables, however reflectance cannot be measured at this angle due to the Schwarzschild objective used to collect the light. The DAC was tilted to a small angle ($\sim 10^0$) to allow the specular reflection of the laser light to pass through the Schwarzschild objective (we did not apply a small angular correction to the reflectance). This orientation, which required careful alignment of the DAC, is less favorable for transmittance because transmitted light suffers multiple attenuation in the sample chamber before being collected. Since we focused this study on the reflectance, we did not carry out a systematic study of transmission.

At temperatures below the transition, the deuterium is still molecular and reflectance is from the W layer. In this case, no significant change in reflectance was observed during the heating laser pulse. Above a certain temperature, an abrupt increase in reflectance was observed during the time of the heating pulse. This was evidence that the deuterium sample had metallized. The raw reflectance was analyzed using the Fresnel equations to extract a value for the reflectance of LMD. The rise in reflectance was very sharp (over a range of less than 100-150 K) consistent with a first-order phase transition. In Table II, we list the P-T values plotted in Fig. 1 for the phase transition based on rising reflectance.

| Pressure | Temperature |
|---|---|
| 142 +/- 2 GPa | 2400 +/- 110 K |
| 149 +/- 2 GPa | 1975-2150 +/- 110 K |
| 169 +/- 2 GPa | 1980 +/- 70 K |

Table SI II. Pressure and temperature values for the transition to LMD based on rising reflectance. The uncertainties in temperature are estimated as arising from systematic error that is much larger than the uncertainty due to the Planckian fit of the data.

The metallic reflectance signals reported here were reproduced at least twice in different heating cycles. Fig. S16 shows the reflectance determined from data collected during 3 different heating cycles at ~142 GPa; these are in excellent agreement, considering the possible variance of ~50-100 K from run to run and slight changes of pressure due to relaxation over time. Of these, data from run1 for 514 and 980 nm were collected simultaneously; the rest were collected separately. Following each heating cycle, we checked the reflectance at a temperature below the transition temperature, and no change in reflectance was observed. This ensured that the high reflectance was not due to a chemical modification of the sample.

In previous work on hydrogen, the reflectivity of LMH was observed to saturate above a given temperature. This was evidence that we obtained a thick layer of LMH, as transmission was indistinguishable from zero in this region; this saturated reflectivity is the bulk value. In those experiments the sample was heated substantially above the plateau to thicken the sample. In this work, although high reflectivity was observed, we could not saturate the reflectance. At the substantially higher temperatures required to metallize deuterium, the thermal conditions are quite different and we could not produce a large increase in temperature beyond the reflectance threshold (see Fig. SI4) and thus could not sufficiently thicken the LMH. It is possible that this may be overcome by going to higher pressures, where the PPT is expected to occur at lower temperatures. However, achieving higher pressures on deuterium has proven to be challenging.

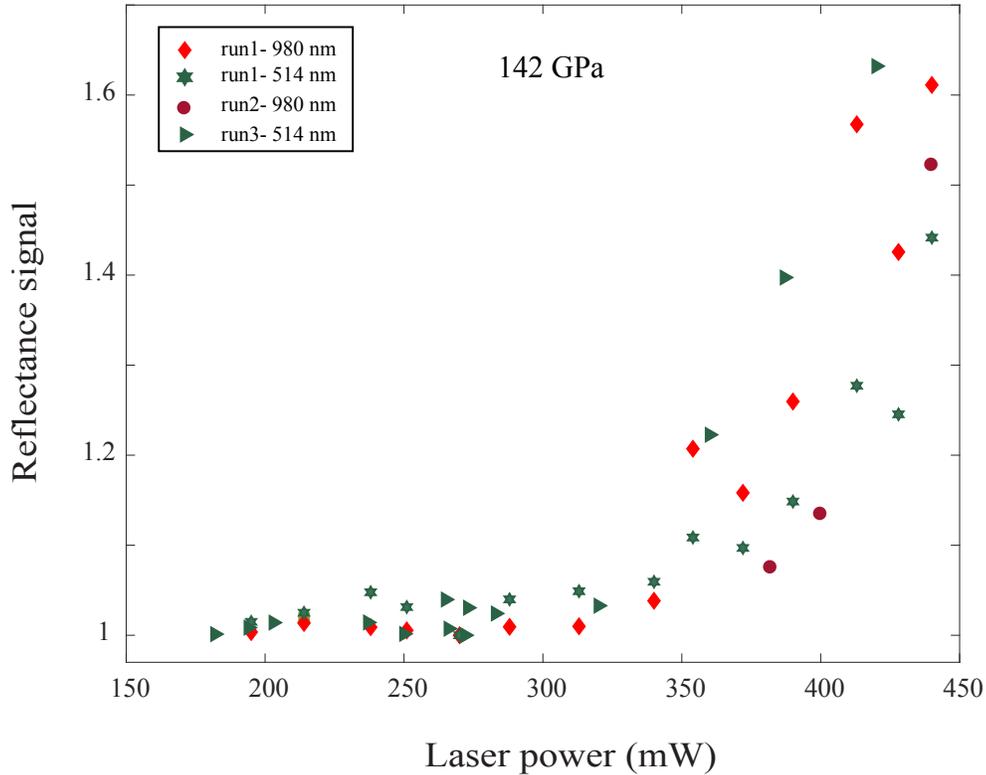

Figure SI6: Reflectance of LMD as a function of average laser power for data collected over three different heating cycles at ~142 GPa.

**Fresnel reflectance analysis**

The analysis of the reflectance data is identical to the analysis performed on hydrogen in Ref. [1], where a detailed description of the procedure is given. The reflectance signal of LMD relative to the W absorber for a given temperature is defined as the intensity of the reflectance signal normalized to the highest-temperature collected below the transition. In the limit of thick LMD layers, the interface between the metallic and molecular deuterium is much smaller than the wavelength of the probe light and thus it can be approximated as a specularly-smooth surface. At about 140 GPa, the index of refraction of solid molecular deuterium is ~2.4[10], and fluid molecular deuterium is estimated to be within a few percent of this value. As the refractive index is similar to that of diamond, we have treated the molecular deuterium and diamond as one interface (Extended Fig. 1). We note that this pressure region for investigating metallic hydrogen and deuterium reflectance properties is convenient, as we do not have to deal with multi-layer interfaces.

**Drude analysis**

In the cases where reflectance was measured for two wavelengths simultaneously, the calculated reflectance of the LMD can be analyzed using the Drude model. The conditions of electron degeneracy (T<< $T_f$, the Fermi temperature), and the characteristic wavelength dependence of the conduction are important criteria for application of the free electron model. At the P-T conditions of these experiments, $T_f = \hbar^2 (3\pi^2 n_e)^{2/3} / 2 m_e k_b = 226{,}091 K$ at 140 GPa, which is large compared to measured transition temperatures of 2000-2500K. $n_e$ was taken as the atom density. Close to the transition region, where reflectance changes abruptly with laser power and we measure low reflectance values, we find that the energy dependence of the optical data cannot be fit to the Drude model. This is indicative of non-free electron like density of states at the metallization phase boundary, also predicted by theory[11-13]. At higher temperatures where we get a thicker metallic layer and higher reflectances, the optical data could be well described by the Drude model (Fig. 3).

We contrast our measured reflectance to that expected in the Mott-Ioffe-Regal minimum metallic conductivity limit (with full dissociation) (*3, 4*). This defines the minimum value of conductivity, and corresponds to a mean-free-path equal to the average interatomic distance, a. The minimum metallic conductivity is given by $\sigma_{min} = e^2/3\hbar a$, where $\hbar$ is the reduced Planck's constant, $a \sim (n_i)^{-1/3}$ is the interparticle spacing, and $n_i$ is the ion number density[14]. The expected reflectance in this limit is plotted against the data in Fig. 3, and clearly cannot account for the high values of reflectance observed in this work.

**Estimates of the Magnitude of the Isotope effect**

Below we make an estimate of the isotope effect based on a statistical model that includes the effect of ZPE and thermally populated vibrational states. The vibrational energies, G, of an isolated diatomic molecule relative to potential energy minimum can be described for a harmonic oscillator as[15]

$$G(v) = \omega_e \left(v + \frac{1}{2}\right) - \delta_e (v + 1/2)^2$$

Here, $v$ is the vibrational quantum number, $\omega_e$ is the vibrational harmonic frequency and $\delta_e$ is the anharmonicity constant.

The zero-point energy, ZPE, is then equal to $G(0) = \frac{1}{2}\omega_e - \frac{1}{4}\delta_e$. In hydrogen or deuterium, $\omega_e$ is just the vibron or the molecular fundamental stretching mode 4401.2 and 3115.5 cm$^{-1}$, respectively. The anharmonic constant is substantially smaller, 121.34 for hydrogen and 61.82 cm$^{-1}$ for deuterium. The isotopic difference in the ZPE is thus

$G_{H2}(0) - G_{D2}(0) \sim 0.079$ eV (916 K), which accounts for the isotopic shift in the dissociation energies of the two free molecules,

Similar order of magnitude analysis could then be carried out in the high-pressure dense phases, where to a first order, the vibrational mode remains the main contributor to the ZPE. At increasing compression, the intramolecular bond weakens as evidenced by the observed decline in the Raman vibron frequency. At 140 GPa, $\omega_{vib}$ is 4070 and 2991 cm$^{-1}$ for hydrogen and deuterium, respectively. $G_H(0) - G_D(0) = 0.067$ eV or 776 K. This value is remarkably close to the observed isotopic shift in the dissociation phase line reported in this manuscript, 700 K.

One can also calculate the ZPE contribution to the molecular partition function (q), as was described in Caillabet et al. (Eqs. 37, 34). The free energy (F) is

$$F_{mol}^{qc}(T) = -K_B T \ln(q_{mol}^{qc})$$

$$q_{mol}^{qc} = \frac{\exp(-\Theta_{vib}/2T)}{1 - \exp(-\Theta_{vib}/2T)} \frac{\Theta_{vib}}{T}$$

At 1300 K (metallization trasition in hdyrogen), $F_{hydrogen}^{qc} - F_{deuterium}^{qc} = 510\ K$.

**Comparison to other experimental studies**

It is interesting to compare our work to previous dynamic and static experiments on dense fluid hydrogen and deuterium. In Fig. SI7, we plot all the relevant experiments that have probed the fluid regime close to the phase line. Each experiment employed different diagnostics and relied on differing criteria to describe the onset of metallic behavior. The onset of reflectance described both in this work and in previous work is also shown. In addition to the onset of reflectance, we plot the onset of absorption in experiments, where it was observed. In the hydrogen experiments the onset of absorption was at the onset of plateaus, whereas the onset of reflectance occurred at temperatures above the plateau (see Fig. 4 of ref. [1]). Note that the P-T points corresponding to the observation of the plateaus[1,16] coincides with the onset of absorption in both static and dynamic experiments. The most recent absorption results from ramp-laser compression experiments, which found metallic reflectance at 200 GPa and above 1000-2000 K, appears to agree with our current results[17,18]. There is some overall consistency between the dynamic and static experiments: as one enters into the metallic region there is first absorption of light and at a higher P or T there is a rapid increase in reflectance, except for a report in the static pressurization case of McWilliams et al[19].

McWilliams et al combined pulsed laser heating with transmission measurements and reported opacity in heated hydrogen[19]. These points are shown as red triangles (H-Abs) in Fig. SI7. The authors state that hydrogen is not metallic, but either semiconducting or semi-metallic. Silvera et al[20] have commented that this experiment lacks the time resolution to detect a transition to LMH and pointed out a serious experimental design flaw which casts doubt on their results.

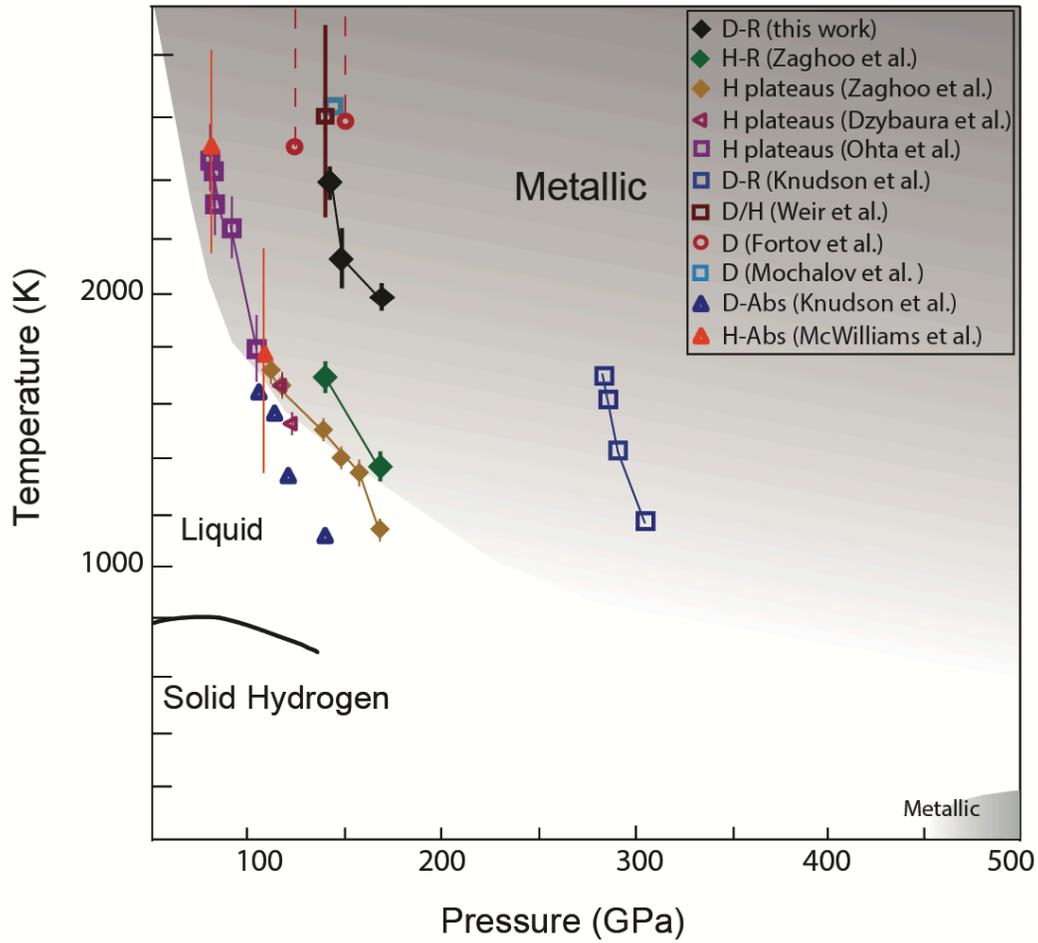

Figure SI7. Phase diagram showing experimental studies for the metallization phase line in hydrogen and deuterium. The onset of absorption in both static[1] and dynamic[9] measurements coincides with the observation of plateaus in heating curves. The onset of reflectance in deuterium from this work is in excellent agreement with previous shock-wave experiments that have observed plateauing in electrical conductivity in fluid hydrogen[21] or a density jump in fluid deuterium[22,23].